\documentclass[a4paper,man,british]{apa6}
\usepackage{amsmath} 
\usepackage{epstopdf}
\usepackage[caption=false]{subfig}
\usepackage{xcolor}
\usepackage[natbibapa,nodoi]{apacite}
\title{Network Mediation Analysis Using Model-based Eigenvalue Decomposition}

\shorttitle{Network analysis}

\threeauthors{Chang Che}{Ick Hoon Jin}{Zhiyong Zhang}

\threeaffiliations{Department of Psychology, University of Notre Dame}{Department of
Applied Statistics, Yonsei University }{Department of Psychology, University of Notre Dame}

\authornote{Correspondance should be sent to Chang Che, 430 Corbett Family Hall, Department of Psychology, University of Notre Dame, IN 46556. Email: cche1@nd.edu}

\abstract{This paper proposes a new two-stage network mediation method based on the use of a latent network approach -- model-based eigenvalue decomposition -- for analyzing social network data with nodal covariates. In the decomposition stage of the observed network, no assumption on the metric of the latent space structure is required. In the mediation stage, the most important eigenvectors of a network are used as mediators. This method further offers an innovative way for controlling for the conditional covariates and it only considers the information left in the network. We demonstrate this approach in a detailed tutorial \textsf{R} code provided for four separate cases -- unconditional and conditional model-based eigenvalue decompositions for either a continuous outcome or a binary outcome -- to show its applicability to empirical network data.}


\begin{document}

\maketitle

\section{Introduction} \label{Introduction}
Network analysis is becoming a popular interdisciplinary research topic in computer science, statistics, sociology, political science, and psychology. For example, two recent special issues in {\em Journal of Research on Adolescence} \citep{veenstra2013network} and {\em Psychosocial Intervention} \citep{maya2015network}, demonstrated the usefulness of network analysis to psychological research. People endeavor to understand how individual similarity and dissimilarity impact their friendship or how social impact and peer influence explain human behaviors like smoking, alcohol use, and drug use. To better investigate such questions, there have been attempts to disentangle the relationship between individual-specific variables and their social network \citep{WassermanPattison1996,pattison1999logit,van2004p2,robins2007,Hunter2008,Hoff2008Multiplicative,HoffAME2018,Liu2018}. These attempts in network analysis are useful in providing a general picture of the social system. However, they are limited in identifying processes or mechanisms underlying the associated links, where we believe network mediation analysis fills the gap naturally.

Mediation and related methods are utilized ubiquitously in psychological theory and research \citep{MacKinnon2007}. The origin of mediation analysis roots from the effort to investigate the mechanism of several variables and possible causality to justify that prevention variables can influence outcome variables \citep{BaronKenny1986}. Although the determination of causality may easily fail without a sounded theoretical base \citep{muller2005moderation,james2008path,stone2008relative,imai2010general,bullock2010yes,imai2011}, mediation analysis and related ideas prevail in psychological studies for years and provide valuable insights on variable relationships. 

Despite the wide applications of mediation methods, little research has been conducted to incorporate network data in a mediation model. This is likely because network data have different format and dimension, which makes the use of network data in mediation analysis very difficult. Nonetheless, it now comes to the point that the use of networks in mediation analysis might be achieved. For example, there are a limited number of studies accomplished the transition from the phase of studying network descriptive statistics to the one of using network characteristics and structure to answer individual-specific research questions. However, we have not seen much research in psychology or other social science disciplines that consolidates network structure with mediation analysis. 

Of the exceptions, \cite{Sweet} proposed to model networks as a mediator through a mixed-membership stochastic block model. Yet, the method views each network as an observation and the summary statistic of the networks as the mediator, and is not suitable for the analysis of one single network. In that method, the mediator can be the density statistics or the parameters of a model in which the network is fitted. \cite{Liu2018} proposed to use coordinates of a Euclidean latent space formed by a network as the mediators in a mediation model. The Euclidean latent space coordinates represent the extent of the similarity of individuals in the observed network. Their model is intended to study the impact of the closeness in networks on outcome variables. However, the assumption of Euclidean latent distance can be easily violated. Furthermore, the interpretation of mediators is extremely hard when the mediators are coordinates but not latent distance, given that the original intention is to use latent similarity for the interpretation of the mediators. 

To contribute to the literature on network mediation analysis, we propose a model based on the use of an alternative latent network approach -- a model-based eigenvalue decomposition method. This approach does not require any assumption on the metric of the latent space structure. In the model, the most important eigenvectors of a network are used as mediators. It also offers an innovative way for controlling the conditional covariate terms to only consider the information left in the network. Therefore, it can be applied to empirical network data and contribute insights to the explanation of social behaviors, especially when there is confounding effect of nodal covariates. The new model also overcomes another challenge of utilizing network structure in mediation analysis -- the dilemma between the urge to simplify the complex observed social network by eliminating the noise and the need to preserve the most valuable information of a network in a justified way. Not only does the model-based eigenvalue decomposition reduce trivial noisy information, but it also preserves dependency structure information in the observed network of various degrees simultaneously \citep{Hoff2008}.

To summarize, the main purpose of this study is to present a new network mediation model based on a model-based eigenvalue decomposition latent network approach and to show how to apply the model in real data analysis using R. The rest of this article is organized as follows. First,  we introduce a real network dataset on college friendship, which is used throughout the entire paper. Then, we briefly review the mediation analysis and the model-based eigenvalue decomposition approach. After that, we show how to combine mediation analysis with the eigenvalue decomposition, including the network mediation models and the related estimation methods. Both continuous and categorical outcomes will be considered. Following it, we show how to analyze the friendship network data step by step using R. Finally, we discuss the limitations and future directions of our study.

\section{Friendship Network Data} \label{FriendshipNetworkDataset}

Throughout the paper, we will use a data set collected by the Lab for Big Data Methodology at the University of Notre Dame. The data include a friendship network of 165 undergraduate college students, in which the information on whether two students are friends or not is available. In addition, basic demographic information such as age and gender as well as self-reported behavioral data on smoking and alcohol use are available.

A network has at least two components -- vertices (nodes, or actors) and edges (links, or ties). A network can be represented using a square matrix, called an adjacency matrix or sociomatrix, denoted as $\mathbf{A}$ in this article. In social networks, the network nodes are often people and the edges can be any relationship among people such as friendship. A network can be directed or undirected. A directed network is asymmetric where person $i$ may view person $j$ as a friend while not visa versa. For an undirected network, the relationship is mutual where person $i$ and person $j$ view each other as friends. There are binary and valued networks. For a binary network, only 0 and 1 are used and a value 1 indicates the existence of an edge. For a valued network, the edges can take different values to represent the strength of the relationship.

\begin{figure}[h]
\centering
\subfloat[Network plot.]{%
\resizebox*{5cm}{!}{\includegraphics{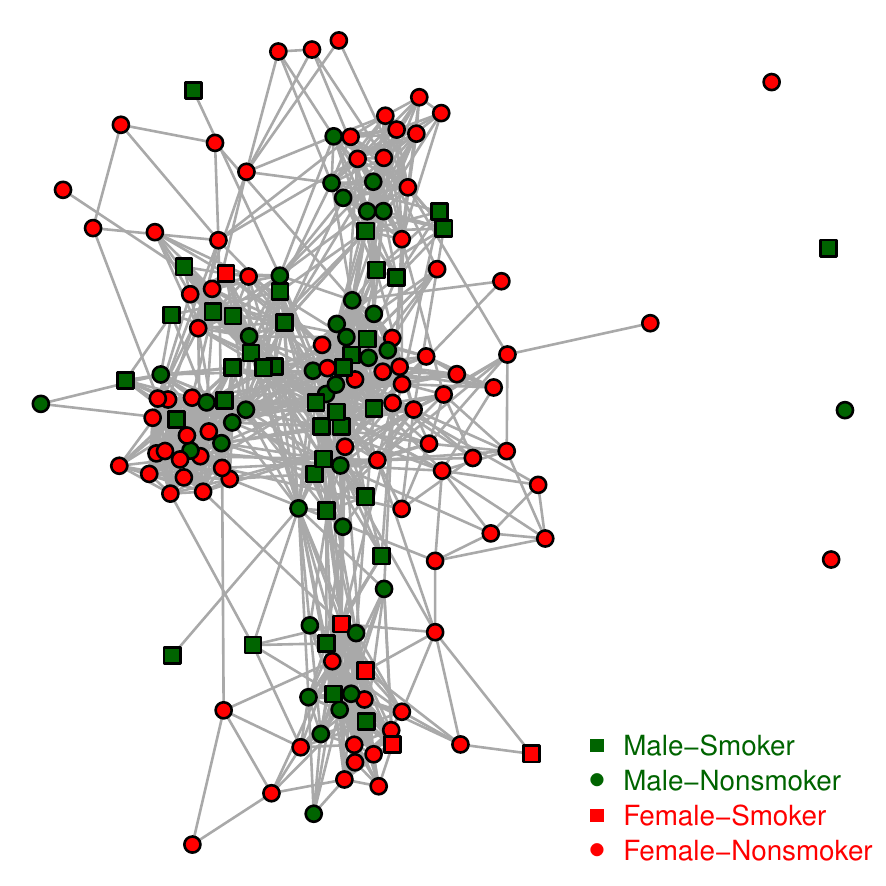}}}\hspace{5pt}
\subfloat[Heatmap of the adjacency matrix. \label{fig:heat}]{%
\resizebox*{5cm}{!}{\includegraphics{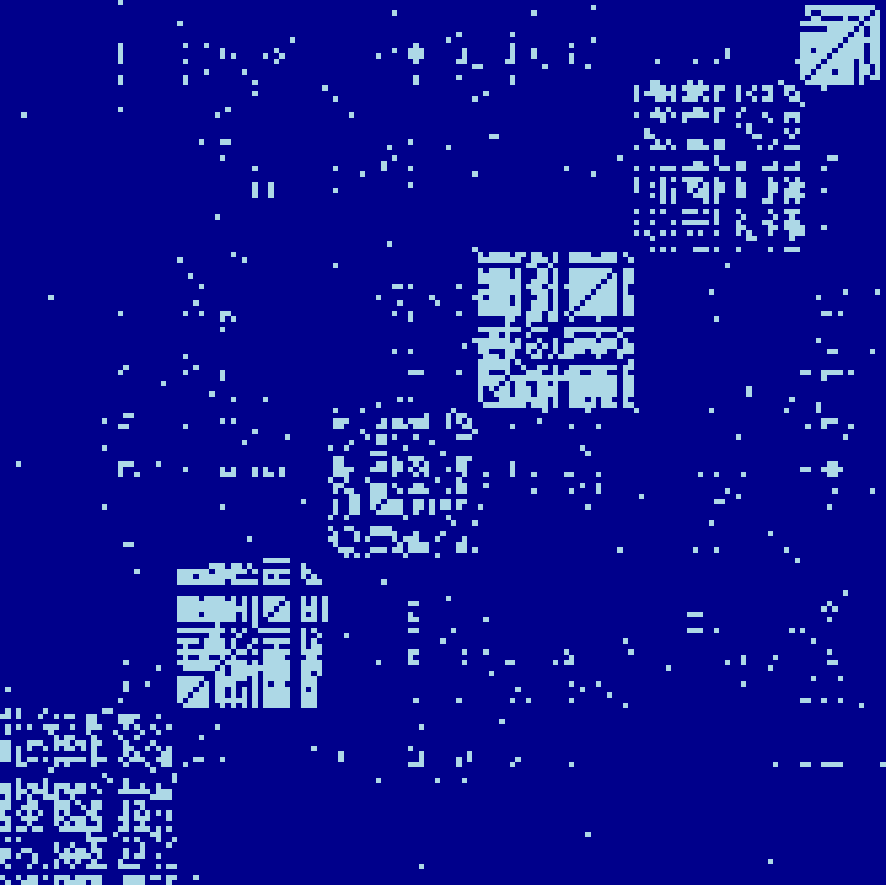}}}
\caption{Visualization of the adjacency matrix of the friendship network data.} \label{fig:netadj}
\end{figure}

Figure \ref{fig:netadj} depicts the friendship network data. In Figure \ref{fig:netadj}(a), each square or circle represents a student. If two students are friends, they are connected by a line between them. If a student smokes cigarettes, the node is a square, otherwise, a circle. The red color presents a female student and the green color male. Figure \ref{fig:netadj}(b) visualizes the adjacency matrix using a heatmap in which the lighter blue color flags a mutual friendship between two students. The heatmap shows a six-block structure, which might indicates six potential communities or groups in the data.

The density of this friendship network is 0.065, calculated as the ratio of the observed edges over all possible edges of 165 nodes for this binary network. The diameter, the value of the longest distance between any two students, of this network is 6, corresponding to the Small-world phenomenon \citep{watts1998collective,milgram1967small}. Transitivity coefficient (also called clustering coefficient) measures the tendency of building a third edge among a triangle structure when there are already two edges. The transitivity coefficient of this network is 0.42, implying a high tendency to the closure of the triangle among three students and further confirming the common notion that a friend of a friend can easily become a friend and the existence of dependency in this network.

\section{Mediation Analysis}\label{Mediation Analysis}

\begin{figure}[h]
\centering
\resizebox*{10cm}{!}{\includegraphics{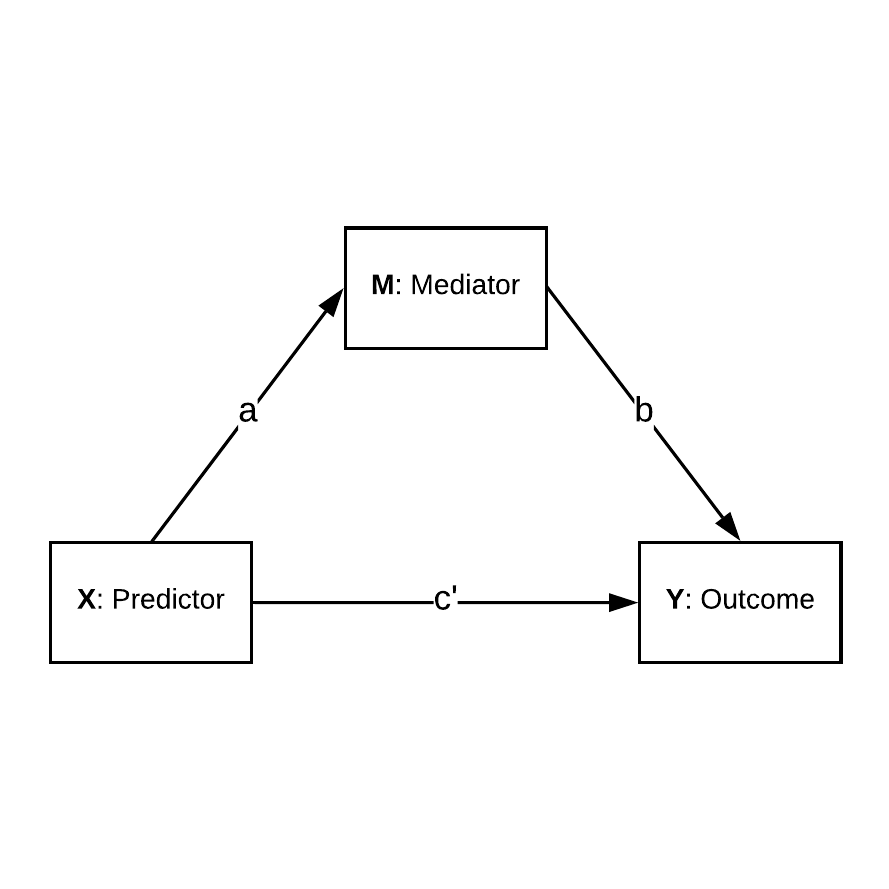}}
\caption{Single mediator regression model.} \label{fig:meddemo}
\end{figure}

Mediation analysis is widely used in social sciences, especially in psychology. Such analysis can evaluate whether a mediating variable transmits the effect from an independent variable onto a dependent variable \citep{BaronKenny1986}. It roots from the reality that numerous research questions in social science suggest a chain of relations. The simplest mediation model is the one with a single mediator, shown in Equation (\ref{eq:med}) \citep{BaronKenny1986}. In Equation (\ref{eq:xy}), $Y$ is the dependent outcome, and $X$ is the independent predictor and $c$ is called the total effect. The mediation model can be written as a combination of the equations in (\ref{eq:regmed}), depicted in Figure \ref{fig:meddemo}. Coefficient $a$ in (\ref{eq:regmed}) represents the relation between $X$ and the mediator $M$. $b$ represents the relation of $M$ to Y adjusted for $X$, $c'$ is the relation of $X$ to $Y$ adjusted for the mediator $M$. There are also terms for the intercepts ($i_1$, $i_2$, and $i_3$), and errors ($\epsilon_1$, $\epsilon_2$, and $\epsilon_3$).

\begin{subequations} \label{eq:med}
\begin{align}
    & Y_i = i_1 + c X_i + \epsilon_{1i}. \label{eq:xy}\\
&\begin{cases}    
    Y_i &= i_2 + c'X_i + b M_i + \epsilon_{2i}, \\
    M_i &= i_3 + aX_i + \epsilon_{3i}. 
\end{cases} \label{eq:regmed}
\end{align}
\end{subequations}

In the model, $ab = a\times b$ is called a mediation effect, or indirect effect, which is equivalently obtained by $c-c'$ in the ordinary least squares (OLS) framework \citep{YuanMacKinnon2009}. The rationale is that mediation depends on the extent to which the predictor $X$ changes the mediator, $a$, and the mediator affects the outcome variable $Y$, $b$. $c'$ is called the direct effect, the direct influence of $X$ on $Y$ controlling for the mediator. $c$, the total effect, is the sum of the direct effect and indirect effect from $X$ to $Y$.

The simple mediation model above can be generalized to an ordinary logistic multiple-mediator model. 

\begin{subequations} \label{eq:logismed}
\begin{align}
Y_i & \sim \text{Bernoulli}(p_i),  \label{eq:bernoulli}\\ 
\text{logit}(p_i) & =  i_2 + c'X_i + b_1 M_{1i} + \dotsc + b_Q M_{Qi},\\
M_{ki} & = i_{3k} + a_k X_i + \epsilon_{3ki},\ k = 1,2,\dotsc,Q.
\label{eq:bernmed} 
\end{align}
\end{subequations}

In the model, the outcome variable $Y$ is binary and follows a Bernoulli distribution with probability $p$, where $p_i = Pr(Y_i = 1)$, the probability for the dichotomous outcome to be 1 (See Equation \ref{eq:bernoulli}). $\text{logit}(p_i) = ln(\frac{p_i}{1-p_i})$ is the natural log of the odds that $Y$ equals 1. With a total of $Q$ mediators, each $M_k, k=1,\ldots,Q$ has its own coefficient $a_k$ of the predictor $X$. The mediation effect for each mediator can be defined as $(ab)_k = a_k \times b_k$ and the total mediation effect is the sum of all the products of $a_k$ and $b_k$ as $\sum_{k=1}^{Q} a_k b_k$, denoted as $ab$.

\section{Model-based Eigenvalue Decomposition for Network Analysis (Eigenmodel)}\label{Model-based Eigenvalue Decomposition}

\subsection{Network Models and Dependency Structure}\label{Network Models and Dependency Structure}

Models have been developed for network data. For example, the random graph or network generating models trace their root back to Erd\"{o}s-R\'{e}nyi model (also known as the classical random graph model,  \citealp{Erdos1960}), where edges are generated randomly with a fixed probability. However, the model is incapable of specifying the correct distribution of the degree of nodes or the clustering tendency in social networks in reality \citep{newman2002,daudin2008}. To better model the clusters in random graphs, the stochastic block structures model (SBM, analogous to the latent class model, \citealp{nowicki2001}) was proposed. It assumes that the possibility of edges between nodes within the same classes is equal and the nodes within a class share the same possibility to build connection with nodes that belong to an outside class \citep{bookKolaczyk}. One innovation of the stochastic block models is the incorporation of latent class variables. There was also a growing interest in exponential random graph models for social networks,  known as the $p*$ class of models \citep{WassermanPattison1996,robins2007}. Exponential random graph models inspire researchers to introduce the concept of nodal homophily and incorporate dyadic covariates, which has promoted the network research in social science disciplines \citep{Hunter2008}.

The models mentioned above to some extent assume the independence among edges in the network. However, the assumption hardly holds in a social network. For instance, in the exponential random graph model framework, the edge residual of person $i$ and $j$ may not be independent of a third person $k$, conceptualized as a ``third-order dependency'' \citep{Hoff2005,WassermanFaust1994}. The network dependence can also arise at different orders \citep{HoffAME2018} -- second-order dependencies, such as degree heterogeneity, within-actor correlation, and reciprocity, and third-order or higher-order dependencies, such as transitivity, balance, and clustering. They link to non-zero second or third order moments of residuals mathematically \citep{HoffAME2018}. Degree heterogeneity means that there is variation in the rows or columns of the adjacency matrix. Also, there may be a covariance/correlation between rows and columns. Reciprocity implies the phenomenon that if you like me then I like you. Transitivity often refers to the phenomenon that a friend of a friend is a friend. Balance suggests that the enemy of my friend is an enemy. Clustering is the phenomenon in which a subset of nodes exhibit a large number of within-group ties and relatively few ties outside of the group. These distinctive features of social networks are related to the ``small world problem'' \citep{milgram1967small,Newman2000}. 

All the dependent structures make the development of network models challenging since independence assumptions are assumed by many statistical models. Also, because of the dependence structure, many deterministic approaches for analyzing binary networks are not appropriate \citep{robins2007}. For instance, social relation model (SRM, \citealp{malloy1986social}) and its extension, social relations regression model (SRRM, \citealp{hoff2015dyadic}) are unable to catch third-order or higher-order dependencies \citep{HoffAME2018}. 

\subsection{Latent Methods for Network Analysis}\label{Latent Methods for Network Analysis}

Latent variable approaches have been proposed to model the dependence features of networks. They are built on the idea that edges may arise at least in part from unmeasured and possibly unknown variables \citep{bookKolaczyk}. Particularly, various latent models have been proposed to represent the observed link connecting node $i$ and node $j$ as a function of node-specific latent variables.  

Among them, the latent class model \citep{nowicki2001} assumes stochastic equivalence where the probability of an edge between two nodes depends only on the latent classes to which they belong. Latent class model takes into account this pattern for symmetric networks, in which the latent unobserved membership accordance or discordance is specified as the latent effects. 

For a binary symmetric adjacency matrix $\mathbf{A}$, the nodes are categorized into $Q$ classes, also referred to as blocks, $i \in \{1,\dots,Q\}$. For an entry $a_{ij}$ of $\mathbf{A}$, the probability of the observed edge of nodes $i$ and $j$ is modeled as a function of the pair relation of the latent classes. The pair relation of node $i$ and $j$ can take values from the unordered set $ \{1,\dots,Q\}^2$ with $\frac{Q(Q+1)}{2}$ possible elements at most. The probability of the pair relation can be summarized in such a finite set,
$$
m=\{\alpha({u}_i,{u}_j)|{u}_i,{u}_j \in {1,\dots,Q} \}
$$ 
\noindent where ${u}_i$ is node $i$'s latent membership. The set $m$ is of the same size of $ \{1,\dots,Q\}^2$ and the symmetric function $\alpha$ is identical for $\alpha({u}_i,{u}_j)$ and $\alpha({u}_j,{u}_i)$. Moreover, for any three distinct nodes $i$, $j$, and $h$, if $i$ and $j$ belong to the same latent class, then $Pr(a_{ih}=1)$ and $Pr(a_{jh}=1)$ are identical because $\alpha({u}_i,{u}_h)$ and $\alpha({u}_j,{u}_h)$ take the same value.

We summarize this simple symmetric case of the latent class models in Equation (\ref{eq:latentclass}).

\begin{equation} \label{eq:latentclass}
\begin{aligned}
    Pr(a_{ij}=1) &= p_{ij}\\
    \delta (p_{ij}) &= \alpha({u}_i,{u}_j )\\
\end{aligned}
\end{equation} 
The function $\delta$ is a link function to the probability of whether the value is one or zero based on the latent probability input $\alpha({u}_i,{u}_j)$. For example, $\delta$ can be an identical function.

The latent class model is straightforward in identifying the similarity of groups of nodes. It has a distinct advantage in detecting network clusters. However, for the nodes falling in between clusters without a strong clustering tendency, this model performs poorly since it assumes that the links are conditionally independent given the latent class membership of each node. 

The latent distance model has been proposed as a popular alternative to the latent class model \citep{Hoff2002}. The model assumes the link of $i$ and $j$ is a function of their latent positions $\mathbf{u}_i$ and $\mathbf{u}_j$. Using a chosen distance measure, the similarity of two nodes, or the across-nodes variation in the latent space can be quantified. The probability of a relational link between two individuals should increase as the characteristics of the individuals become more similar. Consequently, a subset of individuals  with a large number of social ties between them may indicate that there is a group of individuals who have nearby positions in the latent social space. For example, the Euclidean latent space assumes that each node holds a position in a latent Q-dimensional Euclidean space with the number of dimensions Q given. Each axis of the Euclidean space serves as a latent factor influencing the formation of connections between nodes. For nodes $i$ and $j$, $\mathbf{u}_i$ and $\mathbf{u}_j$ represent their latent position vectors, respectively. Thus, the Euclidean distance between the two nodes is $\sqrt{(\mathbf{u}_i -\mathbf{u}_j)^T (\mathbf{u}_i -\mathbf{u}_j)}$. Now we summarize the latent space model in Equation (\ref{eq:latentdist}):

\begin{equation} \label{eq:latentdist}
\begin{aligned}
  Pr(a_{ij}=1) &= p_{ij} \\
  \delta (p_{ij}) &= \alpha(\mathbf{u}_i,\mathbf{u}_j)\\
  &= a - \sqrt{(\mathbf{u}_i -\mathbf{u}_j)^T (\mathbf{u}_i -\mathbf{u}_j)},
\end{aligned}
\end{equation}
\noindent where $a$ is the intercept. As the Euclidean distance decreases, the probability of an edge will increase. The function $\delta$ is a link function, such as a logit function in this case.

\subsection{Latent Eigenmodel}\label{Latent Eigenmodel}

\citet{Hoff2008} further generalized the latent class and the latent distance models into a latent eigenmodel in the spirit of eigenvalue decomposition. Eigenmodel can capture more connectivity patterns -- analogous to dependency structure, than the latent class and the latent distance models, for a given degree of model complexity $Q$ \citep{Goldenberg2009}. The model is given in Equation (\ref{eq:eigenmodel}): 

\begin{equation} \label{eq:eigenmodel}
\begin{aligned}
  Pr(a_{ij}=1) &= \delta(p_{ij})\\
  \delta(p_{ij})&= z_{ij} \\
  z_{ij} &= \beta h_{ij} + \alpha(\mathbf{u}_i,\mathbf{u}_j)\\
  \alpha(\mathbf{u}_i,\mathbf{u}_j) &= \mathbf{u}_i^T \boldsymbol{\Lambda} \mathbf{u}_j,
\end{aligned}
\end{equation}
where $\boldsymbol{\Lambda}$ is a $Q \times Q$ diagonal matrix with the $k$th diagonal value $\lambda_k$,  $k=1,\ldots,Q$. We add an optional dyadic covariate $h_{ij}$ in the function $\delta$ and let $z_{ij}$ represents the latent effect that is the addition of the effect of the dyadic covariate and the latent effect of the eigenvalue decomposition. We use the terminology of dyadic covariate for the variables that reflect the similarity of a pair of nodes on a certain characteristic. In matrix form, the decomposition can be written as  

\begin{equation} \label{eq:eigenmodel_mat}
\begin{aligned}
  \mathbf{Z} &=  \mathbf{H} \beta + \mathbf{U} \boldsymbol{\Lambda} \mathbf{U}^T.
\end{aligned}
\end{equation}

In this article, such a model-based eigenvalue decomposition method is referred to as ``eigenmodel''. As any symmetric matrix can be approximated with a subset of its largest eigenvalues and corresponding eigenvectors, the variation in a sociomatrix $\mathbf{A}$ can be represented through eigenvalue decomposition. Columns of the matrix $\mathbf{U}$, $\mathbf{u}_1$, $\mathbf{u}_2$, $\dots$, and $\mathbf{u}_Q$ are node-specific latent factors. $\mathbf{U}\boldsymbol{\Lambda}\mathbf{U}^T$ represents systematic patterns in the effects. Equation (\ref{eq:eigenmodel_mat}) is a lower rank decomposition of a matrix with node-specific covariates considered. Such a reduced-rank approximation is to represent the main patterns in the data matrix while eliminating the lower-order noise. 

The latent class model can be interpreted in terms of latent blocks, and the latent distance model can be explained in terms of distance. Although the eigenmodel has been criticized for its lack of interpretability \citep{Goldenberg2009} compared to the latent class model or the latent distance model, it helps the understanding of the observed network while reducing noise. \cite{Hoff2008} discussed a plausible interpretation of this latent eigenmodel. Each node $i$ has a vector of unobserved characteristics $\mathbf{u}_i = (u_{i1}, \dotsc, u_{iQ})^T$, and similar values of $u_{i,k}$ and $u_{j,k}$ will contribute positively or negatively to the relationship between $i$ and $j$, depending on whether $\lambda_k$ is greater than 0 or not. The magnitude of eigenvalues unveils the relative importance of the dimensions of the latent space. For example, when one eigenvalue is considerably larger than others, its corresponding dimension is comprehended as the dominating one. \citet{Hoff2008} argues that the model can represent both positive or negative homophily in varying degrees, as well as stochastic equivalence. Furthermore, we argue the interpretation of eigenvectors for network mediation in exploratory analysis can be accomplished in a similar way as exploratory factor analysis.

The eigenmodel has several advantages. As mentioned earlier, it is a generalization of the latent class and latent distance models, capturing  stochastic equivalence in cluster or block structure as well as the gradual change of similarity in terms of distance. Furthermore, it can model different dependency structures at the same time, including higher order dependencies like transitivity. The specification of an eigenmodel neglects the metric of the latent space, leaving no concern of validation of Euclidean or other similarity metrics. Besides, columns of the $\mathbf{U}$ matrix will be orthogonal for they are the eigenvectors of the decomposition.

\section{Network Mediation Models} \label{Method}

We propose to define network mediation models by combining mediation analysis and eigenmodels. From an observed network, we first extract the eigenvector information through the eigenmodel. Then, we treat the eigenvector as potential mediators to carry out the mediation analysis. 

\begin{figure}{h}
\centering
\subfloat[Model 1: From gender to GPA with an unconditional eigenmodel. \label{fig:medmodel1}]{%
\resizebox*{7cm}{!}{\includegraphics{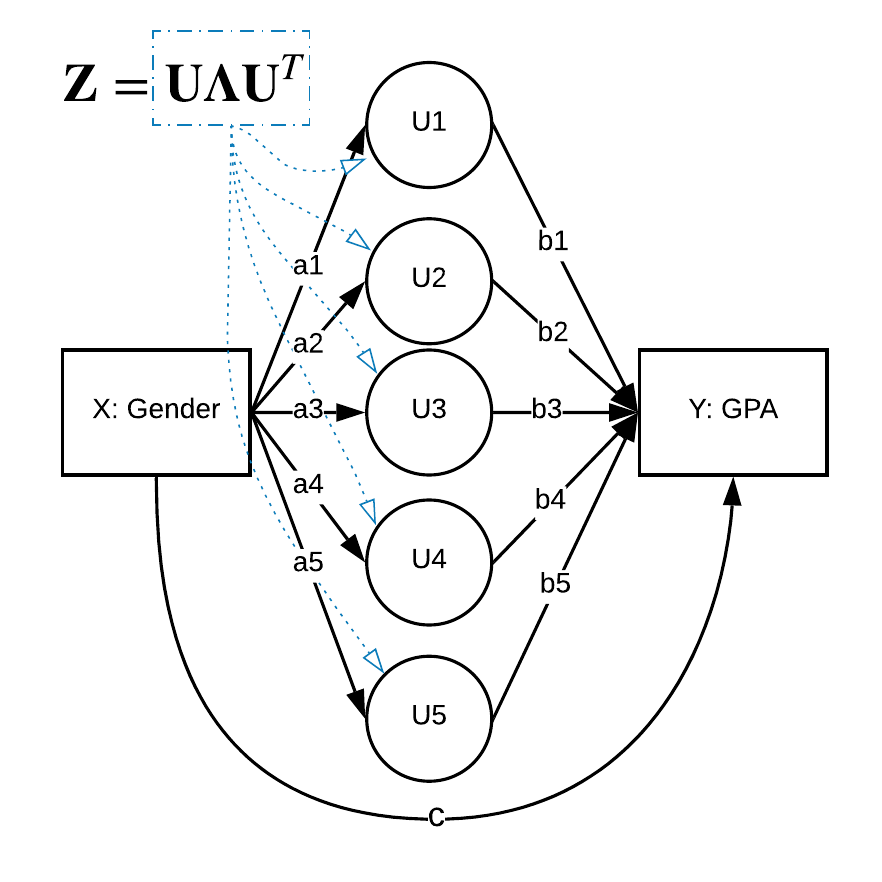}}}\hspace{5pt}
\subfloat[Model 2: From gender to GPA with a conditional eigenmodel. \label{fig:medmodel2}]{%
\resizebox*{7cm}{!}{\includegraphics{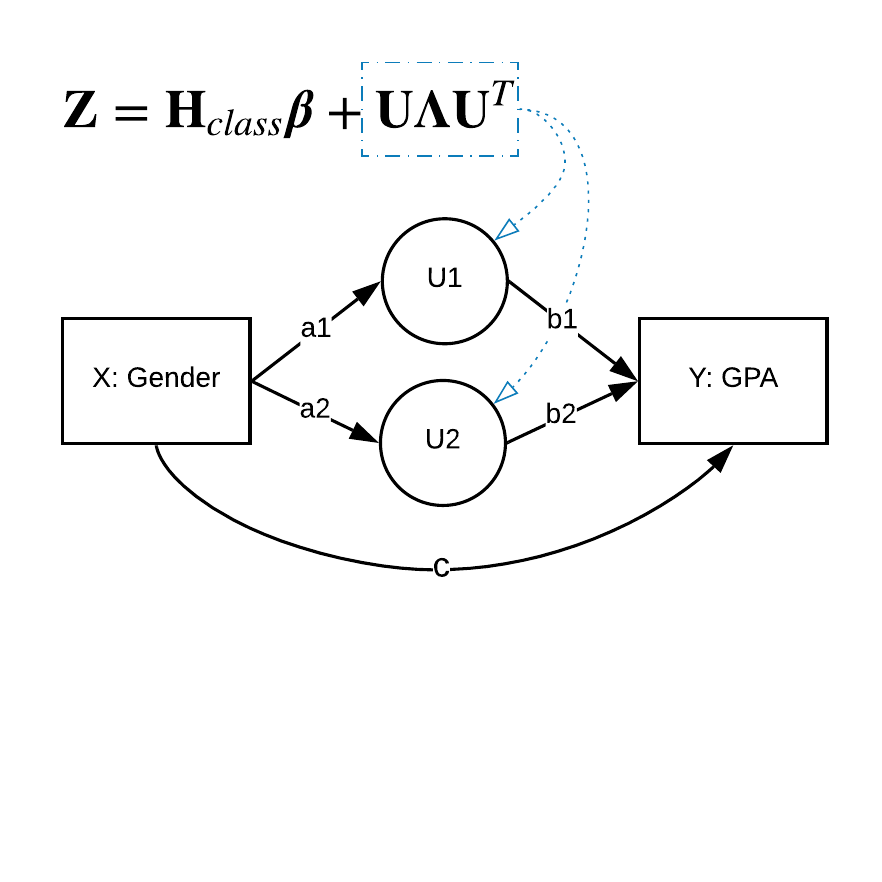}}}

\subfloat[Model 3: From gender to smoking (ifsmoke) with an unconditional eigenmodel. \label{fig:medmodel3}]{%
\resizebox*{7cm}{!}{\includegraphics{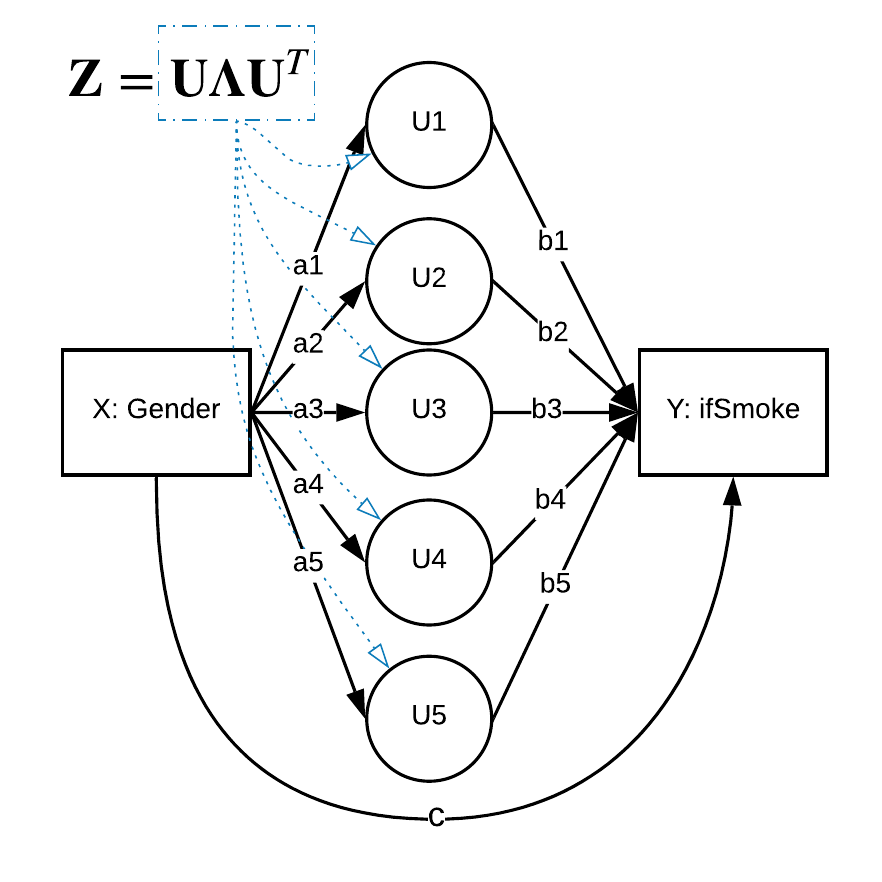}}}\hspace{5pt}
\subfloat[Model 4: From gender to smoking (ifsmoke) with a conditional eigenmodel. \label{fig:medmodel4}]{%
\resizebox*{7cm}{!}{\includegraphics{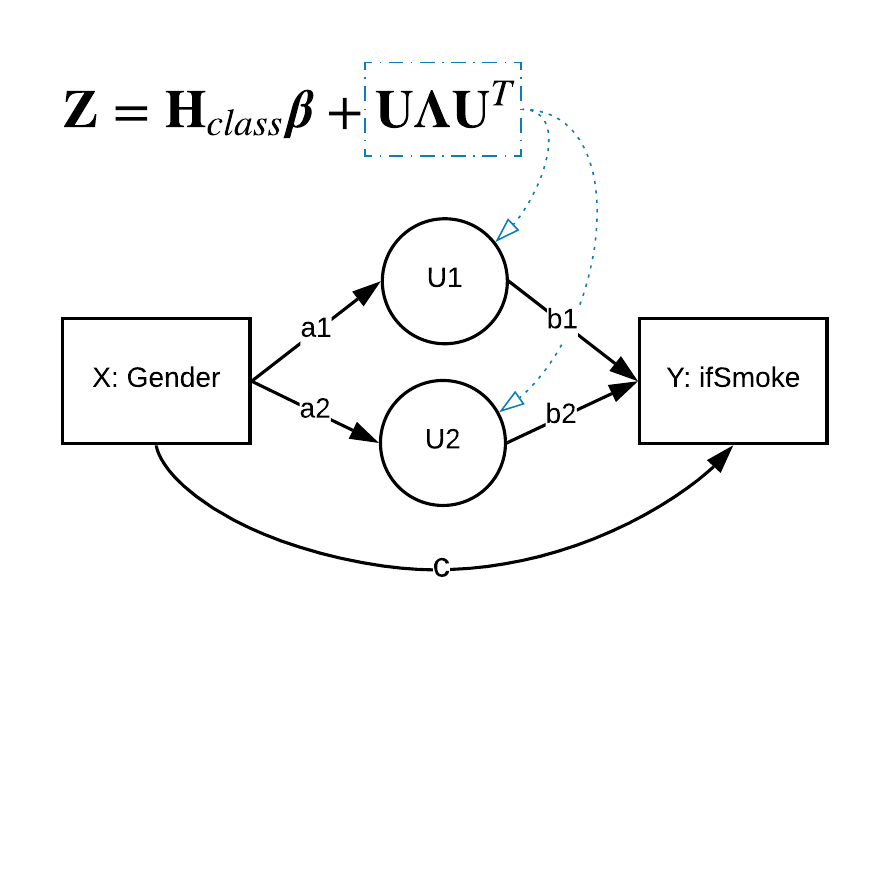}}}
\caption{Four mediation processes in the stage two.} \label{fig:medmodels}
\end{figure}

To concretely introduce the models, we base our discussion on the following practical questions associated with the friendship data: Whether students' academic performance is related to the social network, whether the potential problematic behaviors are related to such a network, and whether a social network's impact can disentangle the mechanism from gender to other outcome variables. We formulate four network mediation models depicted in Figure \ref{fig:medmodels} to provide insights into these questions. Gender is chosen as a natural exogenous predictor, which is not a result of the network. A continuous variable GPA measuring students' academic performance and a binary variable measuring whether a person smokes cigarettes are chosen as the outcome, respectively, to answer whether they are influenced by the social network. All four models in Figure \ref{fig:medmodels} use the latent relational structure extracted from the friendship social network as mediators. The details of the models are discussed below.

\subsection{Network Mediation Model with Continuous Outcome}\label{Regression Mediation}

The network mediation model includes two part. Part one is the eigenmodel that is used to identify the mediators as shown in Equation (\ref{eq:contmodel_eigen}):

\begin{equation} 
\begin{aligned}
 \begin{cases}
  a_{ij} & \sim Bernoulli(p_{ij})\\
  \text{logit}\frac{p_{ij}}{1-p_{ij}} &= z_{ij} \\
  z_{ij} &= \beta \mathbf{h}_{\text{ifsame},ij}  + \mathbf{u}_i \boldsymbol{\Lambda}\mathbf{u}_j^T
  \end{cases}. \label{eq:contmodel_eigen}
\end{aligned}	
\end{equation} 

\noindent Here, $\delta$ is formulated as a logit function linking from the latent variable ${z}$ to the binary observed entry $a_{ij}$ of the adjacency matrix $\mathbf{A}$. 

Part two of the model is the mediation model shown in Equation (\ref{eq:contmodel_med}):

\begin{equation} \label{eq:contmodel_med}
\begin{aligned}
 \begin{cases}
  Y_i &= i_1 + cX_i + \epsilon_{1i}\\
  Y_i &= i_2 + c' X_i + b_1 U_{1i} + \dotsc + b_Q U_{Qi} + \epsilon_{2i}\\
  U_{ki} &=  i_{3k} + a_k X_i + e_{3ki},\  k=1,\dotsc,Q.
  \end{cases}
\end{aligned}	
\end{equation} 

In the model (\ref{eq:contmodel_med}), each mediator $U_k$ is a column vector of the $\mathbf{U}$ matrix. We will use the notation $U$ for mediators for the ease of exposition for the rest of the article. The eigenvalue decomposition should be included and comprehended as a whole in the mediation process  with every eigenvector as a separate mediator. The mediation effect of a network should be calculated as the total mediation indirect effect for all mediators.  A significant indirect effect indicates the evidence for the network to explain the relationship between $X$ to $Y$.

\subsubsection{Unconditional and Conditional Network Model}
We define the unconditional and conditional models in this section. In Equation (\ref{eq:contmodel_eigen}), $z$ represents the latent addition effect of the eigenvalue decomposition of the network and a vector of a conditional covariates $\mathbf{h}_{\text{ifsame}}$. Dyadic covariate $\mathbf{h}_{\text{ifsame}}$ contains $\frac{N(N-1)}{2}$ pieces of pairwise information, whether a pair of nodes is the same on a certain nodal variable. It is a uniform homophily statistic, which will be addressed later. 

The definition of a conditional latent eigenvalue decomposition or unconditional decomposition depends on whether to include a conditional covariates $\mathbf{h}_{\text{ifsame}}$. Without the term $\beta\mathbf{h}_{\text{ifsame}}$, for such an \textbf{unconditional} model, $\mathbf{U} \boldsymbol{\Lambda}\mathbf{U}^T$ is an eigen-space approximation of the observed adjacency matrix. With the term $\beta \mathbf{H}_{\text{ifsame}}$, for such a \textbf{conditional} model, the decomposition $\mathbf{U} \boldsymbol{\Lambda}\mathbf{U}^T$ can be regarded as the remaining information left in the network after controlling for the homophily variable $\mathbf{h}_{\text{ifsame}}$, which is on the condition of the impact of $\mathbf{h}_{\text{ifsame}}$. 

In this article, the matrix $\mathbf{H}_{\text{ifsame}}$ form in Equation (\ref{eq:eigenmodel_mat}) is an $N \times N$ symmetrical matrix for its lower-triangular and upper-triangular entries are both $\mathbf{h}_{\text{ifsame}}$ and diagonals are zero. $\mathbf{H}_{\text{ifsame}}$ is a different approach to arrange homophily in matrix form but has the equivalent information to $\mathbf{h}_{\text{ifsame}}$.

\subsubsection{Determination of $Q$}

The choice of the dimension $Q$ depends on one's own goal of the analysis \citep{Hoff2008}. In general, as the number of dimension $Q$ increases, the information preserved will increase, while ``noise'' contained in the observed network is preserved as well when $Q$ is too large. Although it is not an entirely objective judgment, $Q$ can be determined through a few methods.

For the unconditional model, we can use a similar method in principle component analysis (PCA) -- scree plot \citep{cattell1966scree,lewith2010clinical}, for determining the number of components, to determine the number of eigenvalues here. For this method, we extract the eigenvalues of the observed network directly and then visualize them in a scree plot, where all the positive eigenvalues are in a descendent order. An ``elbow'' in such a scree plot implies the first several important eigenvalues and helps to decide on the dimension of such an eigen-space as the eigenvalues represent the relative importance of the latent eigenvectors.

For the conditional model, exploratory examinations on several dimensions may be necessary. We recommend to try different dimensions in an eigenmodel and extract the eigenvalues. One may start with two latent dimensions as two dimensions are minimal for expanding a reasonable latent space, and stop until there is a negative eigenvalue in the decomposition of the posterior approximation matrix $\mathbf{U} \boldsymbol{\Lambda}\mathbf{U}^T$.

Although the interpretation of negative eigenvalues in the model-based eigenvalue decomposition is not impossible, the accordance of the direction and effect of the mediators are important in the network mediation analysis because the interpretation of the mediation effect shall be considered as a whole package. Therefore, we argue the eigenvectors corresponding to negative eigenvalues should not be included in mediation analysis.

\subsubsection{Nodal Covariates}

In the friendship data, there are variables such as gender and age associated with each of the students. Those variables are called nodal covariates.  \citet{Hunter2008} proposed several ways to transforming nodal covariates to conform to network data: \textit{mean effect} or \textit{absolute difference} for continuous or ordinal variables; and \textit{nodal factor effect} or \textit{homophily effect} for ordinal or categorical variables. A common view of the pair-specific dyadic covariates is that if nodes are similar in terms of a node characteristic, they share similar edge patterns. The uniform homophily statistic is defined by

\begin{equation}
    h(x_i,x_j) = \begin{cases}
        1 & \text{if } i \text{ and } j \text{ both have the same level of the factor}, \\
        0 & \text{otherwise}.
    \end{cases}
\end{equation}

\noindent For instance, if the factor variable $x$ is gender, the uniform homophily is whether two people are of the same gender, and both females or both males will be coded as 1. Hence, a symmetric uniform homophily matrix $\mathbf{H}_{\text{ifsamegender}}$ can be constructed as each entry is either 1 or 0 coded as gender accordance for each pair of people.

For illustration purpose, we employ the uniform homophily statistic defined by \citet{Hunter2008} to construct a dyadic variable based on whether students belong to the same class block in this social network, denoted as $\mathbf{H}_{\text{ifsame}}$ in Equation (\ref{eq:contmodel_eigen}).

\subsection{Network Mediation Model with Binary Outcome}\label{Logistic Mediation}

For binary outcome variable, we need to replace Equation (\ref{eq:contmodel_med}) with Equation (\ref{eq:binmodel}), designed for binary outcome $Y$: 

\begin{equation} \label{eq:binmodel}
\begin{aligned}
\begin{cases}
	Y_i  &\sim Bernoulli(p_i)\\
	logit(p_i) &= i_1 + cX_i\\
	logit(p_i) &= i_2 + c'X_i + b_{1} U_{1i} + \dotsc + b_Q U_{Qi} \\
	U_{ki} &=  i_{3ki} + a_k X_i + e_{3ki},\ k=1,2,\dotsc,Q.
	\end{cases}.
\end{aligned}
\end{equation}

\subsection{Model Estimation} \label{Estimation}

To estimate the model, a two-stage procedure can be used to take advantage of the existing software programs. In stage one, the model-based eigenvalue decomposition is conducted to identify the potential mediators, and in stage two, the mediation effect is estimated and tested. For both stages, Bayesian estimation methods with Markov Chain Monte Carlo (MCMC) can be used. In the current study, the R package \textsf{eigenmodel} is used to estimate the eigenmodel for the network data to get the eigenvectors in the first stage. In the second stage, we use either R package \textsf{blavaan} for continuous outcomes and \textsf{R2jags} for binaray outcomes for mediation analysis.

The \textsf{eigenmodel} package features the functions dedicated for model-based eigenvalue decomposition method. It allows to estimate the $\beta$ coefficient for the dyadic covariate $\mathbf{H}_{\text{ifsame}}$ as well as the decomposed matrix $\mathbf{U} \boldsymbol{\Lambda} \mathbf{U}^T$. The \textsf{blavaan} package \citep{Merkle} is an \textsf{R} package based on \textsf{lavaan} \citep{Rosseel} for estimating Bayesian structural equation models. Since a logistic model under the generalized linear model framework for \textsf{blavaan} is not currently available yet, the package, \textsf{R2jags} \citep{R2jags} is used that provides a relatively easy way for specifying the logistic model.

\section{Examples with Continuous Outcomes}\label{Analysis with Continuous Outcomes}

In this section, we show how to conduct the analysis to investigate whether the friendship network mediates the relationship between gender and academic performance, measured by the continuous outcome GPA.

\subsection{Unconditional Model}\label{Cont Unconditional Model}

We first consider an unconditional eigenmodel without using dyadic covariates. 
First, we use the \textsf{R} package \textsf{eigenmodel} to estimate the eigenmodel for the friendship network data. The \textsf{R} code for the analysis is given below.

\begin{verbatim}
fit1 <- eigenmodel_mcmc(netadj, R=5, S=30000, 
        burn=5000, Nss=(30000-5000))
\end{verbatim}

In the \textsf{eigenmodel\_mcmc} function, network data in the adjacency matrix form should be provided -- here denoted by \textsf{netadj}. Arguments \textsf{S}, \textsf{burn}, and \textsf{Nss} are the length of the whole MCMC chain, the length of the burn-in period, and the number of posterior sample to save. The function also requres a pre-determined number of the latent eigen-space \textsf{R}, which is the number $Q$ in the previous of the article. To determine the number of the latent eigen-space \textsf{R} for this friendship network, a scree plot was used to depict the eigenvalues of the adjacency matrix beforehand. Based both on the ``elbow'' rule and the need for parsimony, a five-eigenvector model was found to be the best in this case.

The convergence of chains can be checked using the \citet{geweke} test in the \textsf{coda} package \citep{coda} . The \textsf{R} code for the convergence diagnosis can be found in the supplementary materials. After achieving convergence, we save the eigenvectors, called \textsf{eigvec1} to be used in stage two, with the code below:

\begin{verbatim}
eigvec1 <- eigen(fit1$ULU_postmean)$vec[,1:5]
\end{verbatim}


Given the 5 eigenvectors identified in the first stage, the mediation model in Figure \ref{fig:medmodel1} (Model 1) is estimated using the \textsf{R} code below:

\begin{verbatim}
Data1 <- data.frame(X = gender, Y = gpa, 
                    U1 = eigvec1[,1], U2 = eigvec1[,2],
                    U3 = eigvec1[,3], U4 = eigvec1[,4],
                    U5 = eigvec1[,5])
med1 <- bsem(model1,data=Data1,sample=30000, 
                      convergence="auto",
                      burnin=5000,n.chains=3)
summary(med1)
\end{verbatim}

Note that in the code, we first organize the input, output and the mediators into a data frame. The function \textsf{bsem} is used to estimate the model. In the function, \textsf{model1} specifies the mediation model as shown in Appendix \ref{append:model1}. In addition, \textsf{data}, \textsf{sample}, \textsf{convergence}, \textsf{burnin}, and \textsf{n.chain} are the arguments for the data set, the length of MCMC chain to take after burnin, to achieve convergence automatically, the length of the burn-in period, and the number of MCMC chains.

The results of the mediation analysis are shown in Table \ref{tab:model1output}, including the posterior estimates (shown as Estimate), standard errors estimates (Post.SD), the 95\% Highest Posterior Density  Intervals (HPD.025, HPD.975). Based on the unconditional eigenmodel, the total mediation effect (\textsf{total}) of the friendship network and the direct effect from gender to GPA \textsf{cp} are both positive and significant, while the mediation/indirect effect \textsf{ab} is negative but not significant. Therefore, the friendship network structure does not seem to mediate the relationship between gender and GPA. 

\renewcommand{\arraystretch}{0.7} 
\begin{table}[!h] 
\caption{The posterior estimates of the mediation coefficients in stage two for Model 1: From gender to GPA with an unconditional eigenmodel.} \label{tab:model1output}
\resizebox{\textwidth}{!}{
\begin{tabular}{rcccc}
\toprule
\multicolumn{1}{l}{Parameter} & Estimate & Post.SD & HPD.025 & HPD.975\\
\midrule
\multicolumn{1}{l}{Direct Effect} \\
  Y $\sim$ X  (cp)  &   0.457 &  0.097 &   0.267 &   0.645  \\
  M1 $\sim$ X (a1) &  -0.005 &  0.017 &  -0.038 &   0.03   \\
  M2 $\sim$ X (a2) &   0.055 &  0.016 &   0.024 &   0.089  \\
  M3 $\sim$ X (a3) &   0.053 &  0.016 &   0.021 &   0.085  \\
  M4 $\sim$ X (a4) &  -0.053 &  0.017 &  -0.086 &  -0.021  \\
  M5 $\sim$ X (a5) &   0.010 &  0.017 &  -0.024 &   0.043  \\
  Y $\sim$ M1 (b1) &   0.535 &  0.477 &  -0.382 &   1.482  \\
  Y $\sim$ M2 (b2) &  -0.217 &  0.557 &  -1.308 &   0.883  \\
  Y $\sim$ M3 (b3) &  -0.141 &  0.556 &  -1.251 &   0.931  \\
  Y $\sim$ M4 (b4) &   0.687 &  0.526 &  -0.353 &   1.698  \\
  Y $\sim$ M5 (b5) &   0.138 &  0.464 &  -0.769 &   1.05   \\
\multicolumn{1}{l}{Mediation Effect} \\
    ab1 & -0.003  &  0.012 &  -0.031 &  0.022\\                  
    ab2 & -0.012  &  0.032 &  -0.077 &  0.054\\    
    ab3 & -0.007  &  0.031 &  -0.072 &  0.055\\   
    ab4 & -0.037  &  0.031 &  -0.103 &  0.021\\  
    ab5 &  0.001  &  0.010 &  -0.017 &  0.024\\   
    ab  & -0.057  &  0.069 &  -0.200 &  0.076\\ 
\multicolumn{1}{l}{Total Effect} \\
    total (c) & 0.400  &  0.076 &   0.250 &  0.547\\
\bottomrule
\end{tabular}
}

\bigskip
\end{table}

\subsection{Conditional Model}\label{Cont Conditional Model}

We now consider a conditional model with a nodal covariate called \textsf{class} indicating whether two students belong to the same class. The following code shows how to construct a conditional model with one dyadic covariate, \textsf{H\_same}. The creation of \textsf{H\_same} follows the same rule of the uniform homophily statistic as previously discussed: The entries of the dyadic covariate matrix \textsf{H\_same} is coded as 1 if two nodes are in the same class, and 0 otherwise. As in Figure \ref{fig:heat}, we spotted a six-block structure, and there are six classes in the friendship data.

The code below estimated the eigenmodel and extracted the eigenvectors. \textsf{R=2} was determined the way as described before -- we fitted all candidate latent dimensions, and found there would be negative eigenvalues after two dimensions. 

\begin{verbatim}
library(eigenmodel)
fit2 <- eigenmodel_mcmc(netadj,X=H_same,R=2,S=30000,
                        burn=5000,Nss=(30000-5000))
eigvec2 <- eigen(fit2$ULU_postmean)$vec[,1:2]                        
\end{verbatim}

The mediation (Model 2) depicted in Figure \ref{fig:medmodel2} is then estimated using the R code below, and the mediation model specification is given in the Appendix \ref{append:model2}. 

\begin{verbatim}
Data2 <- data.frame(X = gender, Y = gpa, 
                    U1 = eigvec2[,1], U2 = eigvec2[,2])
med2 <- bsem(model2,data=Data2,sample=30000,convergence="auto",
             burnin=5000,n.chains=3)
summary(med2)
\end{verbatim}

The results of the analysis is shown Table \ref{tab:model2output}. For the conditional eigenmodel, mediators were the remaining information of the friendship network after the nodal covariate class was controlled. As in the unconditional model, the total effect and the direct effect from gender to GPA are positively significant but the mediation effect \textsf{ab} is  not significant.

\begin{table}[!h]
\caption{The posterior estimates of the mediation coefficients in stage two for Model 2: From gender to GPA with a conditional model.} \label{tab:model2output}
\resizebox{\textwidth}{!}{
\begin{tabular}{rcccc}
\toprule
\multicolumn{1}{l}{Parameter} & Estimate & Post.SD & HPD.025 & HPD.975 \\
\midrule
\multicolumn{1}{l}{Direct Effect} \\
  Y $\sim$ X  (cp)  &   0.303  &  0.111  &  0.090 &  0.526 \\
  M1 $\sim$ X (a1) &  -0.119  &  0.015  & -0.148 & -0.090 \\
  M2 $\sim$ X (a2) &  -0.021  &  0.017  & -0.055 &  0.012 \\
  Y $\sim$ M1 (b1) &  -0.788  &  0.707  & -2.169 &  0.594 \\
  Y $\sim$ M2 (b2) &  -0.177  &  0.464  & -1.113 &  0.701 \\
\multicolumn{1}{l}{Mediation Effect} \\
    ab1 & 0.094  &  0.085  & -0.073  &  0.263 \\  
    ab2 & 0.004  &  0.013  & -0.022  &  0.034 \\    
    ab  & 0.097  &  0.088  & -0.073  &  0.272 \\  
\multicolumn{1}{l}{Total Effect} \\
    total (c) & 0.400 &  0.073  &  0.253  &  0.539 \\
\bottomrule
\end{tabular}
}

\bigskip
\end{table}

\section{Examples with Binary Outcomes}\label{Analysis with Binary Outcomes}

We now show how to conduct the analysis to investigate whether a friendship network to related to gender difference in smoking behavior. To do so, we fit a network mediation model with a binary outcome variable. Although the outcome variable is a binary variable, extracting latent eigenvectors in stage one remains unchanged. Thus, we can use the previous results in stage one. 

For the unconditional eigenmodel in Figure \ref{fig:medmodel3} (Model 3), its specification in \textsf{JAGS} is given in Appendix \ref{append:model3}. In order to estimate the model using the \textsf{R} package \textsf{R2jags}, it is necessary to aggregate the data in a list format and specify the parameters of interest. The \textsf{R} code for the analysis is given below.

\begin{verbatim}
Data3 <- list(N = ncol(netadj), 
               X = gender, Y = ifsmoke, 
               U1 = eigvec1[,1], U2 = eigvec1[,2],
               U3 = eigvec1[,3], U4 = eigvec1[,4],
               U5 = eigvec1[,5])
jags.params1 <- c("total","ab",paste0("ab",1:5),"cp",paste0("a",1:5),
                  paste0("b",0:5),paste0("i0",1:5))
med_3 <- jags(data=Data3, inits=NULL, model.file=model3, 
               parameters.to.save=jags.params1,
               n.chains=3, n.iter=30000)
\end{verbatim}

In \textsf{Data3}, N is the sample size, \textsf{X} and \textsf{Y} represent the predictor gender and the outcome smoking (\textsf{ifsmoke}), and \textsf{U1} to \textsf{U5} are eigenvectors from the eigenmodel. \textsf{jags.params1} lists the parameters of interest. The function \textsf{jags} is used to run the analysis. Although one can specify starting values in the function, we set the initials to be \textsf{NULL} here, which means \textsf{JAGS} will automatically generate the starting values. \textsf{R} function \textsf{jags} also requires the following arguments: \textsf{data} for the dataset, \textsf{model.file} for the model specified (Appendix \ref{append:model3}), \textsf{parameters.to.save} for the parameters of interest, \textsf{n.chains} for the number of MCMC chains to run, and \textsf{n.iter} for the length of each MCMC chain including the burn-in period.

Table \ref{tab:model3output} summarizes the results for the analysis, which has the same format as Table \ref{tab:model2output}. The results show that the  mediation effect of the friendship network is negative but insignificant. The direct effect and the total effect are both negative and significant. 

\begin{table}[!h] 
\caption{The posterior estimates of the mediation coefficients in stage two for Model 3: From gender to smoking with an unconditional model.} \label{tab:model3output}
\resizebox{\textwidth}{!}{
\begin{tabular}{rcccc}
\toprule
\multicolumn{1}{l}{Parameter} & Estimate & Post.SD & HPD.025 & HPD.975 \\
\midrule
\multicolumn{1}{l}{Direct Effect} \\
cp & -2.596 & 0.661 & -3.931 & -1.323\\
a1 & -0.005 & 0.012 & -0.027 & 0.020\\
a2 & 0.056 & 0.012 & 0.036 & 0.077\\
a3 & 0.053 & 0.012 & 0.031 & 0.074\\
a4 & -0.053 & 0.014 & -0.076 & -0.031\\
a5 & 0.010 & 0.013 & -0.014 & 0.032\\
b1 & 1.720 & 3.202 & -4.508 & 7.999\\
b2 & -1.155 & 3.645 & -8.060 & 6.084\\
b3 & -3.486 & 3.760 & -10.698 & 3.737\\
b4 & 9.970 & 3.414 & 3.550 & 16.608\\
b5 & 3.832 & 3.237 & -2.716 & 9.861\\
\multicolumn{1}{l}{Mediation Effect} \\
ab1 & -0.008 & 0.045 & -0.115 & 0.083\\
ab2 & -0.064 & 0.207 & -0.517 & 0.311\\
ab3 & -0.186 & 0.210 & -0.612 & 0.229\\
ab4 & -0.532 & 0.220 & -0.997 & -0.167\\
ab5 & 0.037 & 0.067 & -0.087 & 0.184\\
ab & -0.753 & 0.454 & -1.653 & 0.130\\
\multicolumn{1}{l}{Total Effect} \\
total (c) & -3.349 & 0.604 & -4.492 & -2.182\\
\bottomrule
\end{tabular}
}

\end{table}

The \textsf{R} code for the conditional network mediation model is given below.
\begin{verbatim}
Data4 <- list(N = ncol(netadj),
               X = gender, Y = ifsmoke, 
               U1 = eigvec2[,1], U2 = eigvec2[,2])
jags.params2 <- c("total","ab",paste0("ab",1:2),"cp",
                paste0("a",1:2),paste0("b",0:2),paste0("i0",1:2))
med_4 <- jags(data=Data4, inits=NULL, model.file=model4, 
               parameters.to.save=jags.params2,
               n.chains=3,n.iter =30000)
\end{verbatim}
The \textsf{JAGS} code for the mediation model is given in Appendix \ref{append:model4}. The results from the analysis are summarized in Table \ref{tab:model4output}.

\begin{table}[h]
\caption{The posterior estimates of the mediation coefficients in stage two for Model 4: From gender to smoking with a conditional model.}\label{tab:model4output}
\resizebox{\textwidth}{!}{
\begin{tabular}{rcccc}
\toprule
\multicolumn{1}{l}{Parameter} & Estimate & Post. SD & HPD .025 & HPD .975 \\
\midrule
\multicolumn{1}{l}{Direct Effect} \\
cp & -2.015 & 0.726 & -3.424 & -0.624\\
a1 & -0.119 & 0.010 & -0.134 & -0.103\\
a2 & -0.021 & 0.013 & -0.045 & 0.001\\
b1 & 9.722 & 4.414 & 0.672 & 17.850\\
b2 & -0.274 & 3.785 & -7.392 & 7.237\\
\multicolumn{1}{l}{Mediation Effect} \\
ab1 & -1.153 & 0.531 & -2.145 & -0.078\\
ab2 & 0.005 & 0.092 & -0.176 & 0.211\\
ab & -1.148 & 0.561 & -2.262 & -0.100\\
\multicolumn{1}{l}{Total Effect} \\
total (c) & -3.162 & 0.559 & -4.224 & -2.081\\
\bottomrule
\end{tabular}
}

\bigskip
\end{table}

For the conditional model, when the nodal covariate class is controlled in the eigenmodel stage, we can see that the mediation effect $ab$ is negative and significant. The direct effect and total effect are also negative and significant. It implies the remainder of the network structure after controlling for the class block structure can play an important role in the dynamic between the gender variable and whether or not one student smokes, which functions as a partial mediation. Their friendship with other smokers might strengthen the phenomenon that males tend to smoke, and such an effect can be explained through the mechanism of the network structure.

\section{Conclusion} \label{Conclusion}

This paper developed a network mediation model to incorporate network data into mediation analysis. Given the popularity of mediation analysis, we believe network mediation analysis can provide a unique perspective and additional information to understand social behaviors. Based on the results from the network mediation analysis, we find smoking behavior can be mediated by friendship networks among college students, possibly also generalized to young adults, implying that controlling for peer influence can be a significant potential intervention method. The empirical examples of this article have further shown the usefulness of network analysis in behavioral science.

We argue the model-based eigenvalue decomposition can have great potentials in empirical applications. The eigenvalue decomposition approach effectively extracted information of the observed social network. It has proved to generalize the latent distance model and the latent class model \citep{Hoff2008}. It is an effective approach to approximate the complicated observed network structure with most information recovered by the latent structure. 

The innovative model-based eigenvalue decomposition with controlled dyadic variables, referred to as the conditional model in this article, can control for the effects of nodal covariates. Therefore, researchers can examine the social effect from the remainder of the network information, which can suit many behavioral research interests.

In the process to determine the number of latent dimension of the model-based eigenvalue decomposition, approaches to obtain the optimal number still require further evaluation in the future. Misspecification of the latent dimension and its consequence should be evaluated formally. In addition, although it is hard to interpret the network mediator separately for each latent space, we can gain knowledge about the  impact of a network on behavioral variables as a whole.

The eigenvalue decomposition of networks can be easily extended to singular-value decomposition \citep{Hoff2008,Hoff2008Multiplicative,HoffAME2018} for directed asymmetric networks to obtain information about the idiosyncrasy of nodes and the directions of edges. Applying such a decomposition on directed networks may interest researches with respect to different node characteristics, for example, different levels of sociability and popularity of people in social networks. As symmetric social networks may ignore the possibility for the heterogeneity. Future direction may focus on the illustration of the singular decomposition for asymmetric networks. 

Some node-specific descriptive statistics of the observed network can be included in the mediation process as supplementary information of the observed network. For instance, node degree is a node-specific statistic, and a person's popularity in an undirected network can be represented as node degree. These node features can be a covariate of interest in the social network mediation analysis when the researchers care about idiosyncrasy on popularity.

Moreover, we can extend this model to longitudinal network mediation, as the dynamic changes of social network offer researchers the opportunity to investigate causal effect of the network mediation. The eigenvalue decomposition provides a tool to compare the dynamic changes contained in the observed networks. Thus, causal mediation eventually can be feasible in social network studies.

\bibliographystyle{apacite}
\bibliography{eigenmed}

\appendix

\section{Model 1: Continuous outcome GPA with unconditional eigenmodel}\label{append:model1}

Model 1: 
\begin{equation}
\begin{aligned}
  \text{GPA}_i &= i_1 + c\text{Gender}_i + \epsilon_{1i}\\
  \text{GPA}_i &= i_2 + cp \text{Gender}_i + b_1 U_{1i} + b_2 U_{2i} + b_3 U_{3i} + b_4 U_{4i} + b_5 U_{5i} + \epsilon_{2i}\\
  U_{1i} &=  i_{31} + a_1 \text{Gender}_i + e_{31i}\\
  U_{2i} &=  i_{32} + a_2 \text{Gender}_i + e_{32i}\\
  U_{3i} &=  i_{33} + a_3 \text{Gender}_i + e_{33i}\\
  U_{4i} &=  i_{34} + a_4 \text{Gender}_i + e_{34i}\\
  U_{5i} &=  i_{35} + a_5 \text{Gender}_i + e_{35i}.
\end{aligned}	
\end{equation} 

When the outcome variable is a continuous variable, the analysis using \textsf{blavaan} package is recommended. For \textsf{blavaan} or \textsf{lavaan}, models should be specified as character objects in \textsf{R}. The notation tilda, \verb ~  is the regression operator. Refer to the \textsf{lavaan} manual for further details.

Model 1 is a mediation model with 5 latent eigenvectors as mediators. \textsf{model1} corresponds to Figure \ref{fig:medmodel1}. \textsf{Y}, \textsf{U} and \textsf{X} denote data vectors for the dependent variable \textsf{GPA}, the mediating variable \textsf{U} and the independent variable \textsf{Gender}, respectively. Priors are not specified in the model part for \textsf{blavaan} as we use the default priors for simplicity.

\textsf{R} code for specifying \textsf{model1} for \textsf{blavaan} in the stage two is below:

{\small
\begin{verbatim}

model1 <- ' 
# direct effect:
# from X to Y
Y ~ cp*X 

# mediators:
# from X to mediator 1
U1 ~ a1*X 
# from X to mediator 2
U2 ~ a2*X 
# from X to mediator 3
U3 ~ a3*X 
# from X to mediator 4
U4 ~ a4*X 
# from X to mediator 5
U5 ~ a5*X 
# from mediators to Y
Y ~ b1*U1 + b2*U2 + b3*U3 + b4*U4 + b5*U5 

# define effects
# indirect effect (a*b)
# indirect effect of mediator 1
ab1 := a1*b1 
# indirect effect of mediator 2
ab2 := a2*b2 
# indirect effect of mediator 3
ab3 := a3*b3 
# indirect effect of mediator 4
ab4 := a4*b4 
# indirect effect of mediator 5
ab5 := a5*b5 
# total indirect effect
ab := a1*b1 + a2*b2 + a3*b3 + a4*b4 + a5*b5
# total effect
total := cp + (a1*b1 + a2*b2 + a3*b3 + a4*b4 + a5*b5)

'
\end{verbatim}
}


\section{Model 2: Continuous outcome GPA and conditional eigenmodel}\label{append:model2}

Model 2: 
\begin{equation}
\begin{aligned}
  \text{GPA}_i &= i_1 + c\text{Gender}_i + \epsilon_{1i}\\
  \text{GPA}_i &= i_2 + cp \text{Gender}_i + b_1 U_{1i} + b_2 U_{2i} + \epsilon_{2i}\\
  U_{1i} &=  i_{31} + a_1 \text{Gender}_i + e_{31i}\\
  U_{2i} &=  i_{32} + a_2 \text{Gender}_i + e_{32i}.
\end{aligned}	
\end{equation} 

Model 2 is a mediation model with 2 latent eigenvectors as mediators. \textsf{model2} corresponds to Figure \ref{fig:medmodel2}. \textsf{Y}, \textsf{U} and \textsf{X} denote data vectors for the dependent variable \textsf{GPA}, the mediating variable \textsf{U} and the independent variable \textsf{Gender}, respectively. Priors are not specified in the model part for \textsf{blavaan} as we use the default priors for simplicity reason.

\textsf{R} code for specifying \textsf{model2} for \textsf{blavaan} in the stage two is below:

{\small
\begin{verbatim}
model2 <- ' 
# direct effect
# from X to Y
Y ~ cp*X

# mediator
# from X to mediator 1
U1 ~ a1*X
# from X to mediator 2
U2 ~ a2*X
# from mediators to Y
Y ~ b1*U1 + b2*U2 

# define effects
# indirect effect (a*b)
# indirect effect of mediator 1
ab1 := a1*b1
# indirect effect of mediator 2
ab2 := a2*b2
# total indirect effect
ab := a1*b1 + a2*b2 
# total effect
total := cp + (a1*b1 + a2*b2)

'
\end{verbatim}
}

\section{Model 3: Binary outcome smoking (if-smoke) and unconditional eigenmodel}\label{append:model3}

Model 3: 
\begin{equation}
\begin{aligned}
\text{ifSmoke}_i  &\sim Bernoulli(p_i)\\
logit(p_i) &= i_1 + c\text{Gender}_i\\
logit(p_i) &= i_2 + cp\text{Gender}_i + b_{1} U_{1i} + b_2 U_{2i} + b_{3} U_{3i} + b_4 U_{4i} + b_{5} U_{5i}\\
U_{1i} &=  i_{31i} + a_1 \text{Gender}_i + e_{31i}\\
U_{2i} &=  i_{32i} + a_2 \text{Gender}_i + e_{32i}\\
U_{3i} &=  i_{33i} + a_3 \text{Gender}_i + e_{33i}\\
U_{4i} &=  i_{34i} + a_4 \text{Gender}_i + e_{34i}\\
U_{5i} &=  i_{35i} + a_5 \text{Gender}_i + e_{35i}
\end{aligned}
\end{equation}

When the outcome variable is a binary categorical variable, the analysis using \textsf{R2jags} package is recommended, for \textsf{blavaan} package cannot handle categorical outcomes yet.

\textsf{R} code for specifying \textsf{model3} as \textsf{JAGS} models in \textsf{R} is below, corresponding to Figure \ref{fig:medmodel3}. \textsf{Y[i]}, \textsf{U[i]} and \textsf{X[i]} denote data vectors for the dependent variable smoking (\textsf{ifSmoke}), the mediating variables \textsf{U}'s and the independent variable \textsf{Gender}, respectively. N is the number of observations.

It is worth noting that in the \textsf{R} user interface of \textsf{JAGS} (or \textsf{BUGS}), models should be specified as function objects. The JAGS language, notation tilda, \verb ~  means priors or likelihoods following some distribution; and notation \verb <-  or equal sign is for value assignments. Refer to the JAGS manual for further details.

{\small
\begin{verbatim}

model3 <- function(){
  for(i in 1:N)
  {
    # specify the mediation model U_* = i0_* + a_* X + e_*
    # dnorm(mu, sigma) denotes a normal distribution
    # with the mean mu and precision of sigma (or a variance of 1/sigma).
    U1[i] ~ dnorm(mean.u1[i], prec.u1)
    U2[i] ~ dnorm(mean.u2[i], prec.u2)
    U3[i] ~ dnorm(mean.u3[i], prec.u3)
    U4[i] ~ dnorm(mean.u4[i], prec.u4)
    U5[i] ~ dnorm(mean.u5[i], prec.u5)
    mean.u1[i] <- i01 + a1*X[i]
    mean.u2[i] <- i02 + a2*X[i]
    mean.u3[i] <- i03 + a3*X[i]
    mean.u4[i] <- i04 + a4*X[i]
    mean.u5[i] <- i05 + a5*X[i]
    
    # specify the mediation model 
    # for binary outcome, the logistic model:
    Y[i] ~ dbern(p[i])
    logit(p[i]) <- b0 + b1*U1[i] + b2*U2[i] + b3*U3[i] + b4*U4[i] + b5*U5[i] + cp*X[i]
  }
  # normal priors for coefficients. Huge variances 10^6, essentially noninformative.
  i01 ~ dnorm(0, 1.0E-6) 
  i02 ~ dnorm(0, 1.0E-6) 
  i03 ~ dnorm(0, 1.0E-6)
  i04 ~ dnorm(0, 1.0E-6) 
  i05 ~ dnorm(0, 1.0E-6) 
  b0 ~ dnorm(0, 1.0E-6) 
  a1 ~ dnorm(0, 1.0E-6)
  b1 ~ dnorm(0, 1.0E-6)
  a2 ~ dnorm(0, 1.0E-6)
  b2 ~ dnorm(0, 1.0E-6)
  a3 ~ dnorm(0, 1.0E-6)
  b3 ~ dnorm(0, 1.0E-6)
  a4 ~ dnorm(0, 1.0E-6)
  b4 ~ dnorm(0, 1.0E-6)
  a5 ~ dnorm(0, 1.0E-6)
  b5 ~ dnorm(0, 1.0E-6)
  cp ~ dnorm(0, 1.0E-6) 
  # priors for precisions
  # dgamma (a, b) is a gamma distribution 
  # with the shape parameter a and
  # inverse scale parameter b.
  prec.u1 ~ dgamma(0.001, 0.001)
  prec.u2 ~ dgamma(0.001, 0.001)
  prec.u3 ~ dgamma(0.001, 0.001)
  prec.u4 ~ dgamma(0.001, 0.001)
  prec.u5 ~ dgamma(0.001, 0.001)
  # define the mediated effect
  # indirect effect (a*b)
  ab1 <- a1*b1
  ab2 <- a2*b2
  ab3 <- a3*b3
  ab4 <- a4*b4
  ab5 <- a5*b5
  ab <- a1*b1 + a2*b2 + a3*b3 + a4*b4 + a5*b5
  # total effect
  total <- cp + (a1*b1 + a2*b2 + a3*b3 + a4*b4 + a5*b5)
}
\end{verbatim}
}

\section{Model 4: Binary outcome smoking (ifsmoke) and conditional eigenmodel}\label{append:model4}

Model 4: 
\begin{equation*}
\begin{aligned}
\text{ifSmoke}_i  &\sim Bernoulli(p_i)\\
logit(p_i) &= i_1 + c\text{Gender}_i\\
logit(p_i) &= i_2 + cp\text{Gender}_i + b_{1} U_{1i} + b_2 U_{2i} \\
U_{1i} &=  i_{31i} + a_1 \text{Gender}_i + e_{31i}\\
U_{2i} &=  i_{32i} + a_2 \text{Gender}_i + e_{32i}\\
\end{aligned}
\end{equation*}

R code for specifying \textsf{model4} in the stage two is below, corresponding to Figure \ref{fig:medmodel4}. \textsf{Y[i]}, \textsf{U[i]} and \textsf{X[i]} denote data vectors for the dependent variable \textsf{ifSmoke}, the mediating variables \textsf{U}'s and the independent variable \textsf{Gender}, respectively. N is the number of observations.
  
{\small
\begin{verbatim}
model4 <- function(){
  for(i in 1:N)
  {
    # specify the mediation model U_* = i0_* + a_* X + e_*
    # dnorm(mu, sigma) denotes a normal distribution
    # with the mean and precision of sigma (or a variance of 1/sigma).
    U1[i] ~ dnorm(mean.u1[i], prec.u1)
    U2[i] ~ dnorm(mean.u2[i], prec.u2)
    mean.u1[i] <- i01 + a1*X[i]
    mean.u2[i] <- i02 + a2*X[i]
    
    # specify the mediation model
    # for binary outcome, a logistic model
    Y[i] ~ dbern(p[i])
    logit(p[i]) <- b0 + b1*U1[i] + b2*U2[i] + cp*X[i]
  }
  # normal priors for coefficients. Huge variances, essentially noninformative.
  i01 ~ dnorm(0, 1.0E-6)
  i02 ~ dnorm(0, 1.0E-6)
  b0 ~ dnorm(0, 1.0E-6)
  a1 ~ dnorm(0, 1.0E-6) 
  b1 ~ dnorm(0, 1.0E-6) 
  a2 ~ dnorm(0, 1.0E-6) 
  b2 ~ dnorm(0, 1.0E-6) 
  cp ~ dnorm(0, 1.0E-6) 
  # gamma priors for precisions
  # dgamma (a, b) is a gamma distribution 
  # with the shape parameter a and
  # the inverse scale parameter b
  prec.u1 ~ dgamma(0.001, 0.001)
  prec.u2 ~ dgamma(0.001, 0.001)
  # define the mediated effect
  # indirect effect (a*b)
  ab1 <- a1*b1
  ab2 <- a2*b2
  ab <- a1*b1 + a2*b2 # total indirect effect
  # total effect
  total <- cp + (a1*b1 + a2*b2)
}
\end{verbatim}
}

\end{document}